\documentclass[final] {aipproc}
\usepackage{psfig}
\usepackage{graphicx}
\layoutstyle{6x9}

\newcommand{\lsim}{{\, \lower2truept\hbox{
${< \atop\hbox{\raise4truept\hbox{$\sim$}}}$}\,}}
\newcommand{\gsim}{{\, \lower2truept\hbox{
${> \atop\hbox{\raise4truept\hbox{$\sim$}}}$}\,}}

\begin{document}
\title{Nova Populations}

\author{Massimo Della Valle}{
  address={Osservatorio Astrofisico di Arcetri, Largo E. Fermi 5, 50125 FI, Italy}
}

\begin{abstract}
In this article we review the current status of the stellar population
assignment for novae. Observations in the Milky Way
and in external galaxies point out the existence of two nova
populations: fast and bright novae, mainly originated from massive white 
dwarfs and associated with the {\sl thin disk/spiral arm} stellar population,
and slow and faint novae, originated from lighter white dwarfs and
associated with {\sl thick-disk/bulge} population.
\end{abstract}

\maketitle


\section{Introduction}

Baade \cite{baa44},\cite{baa57} introduced the concept that
different kinds of stellar populations have different spatial
distribution within galaxies (see also Oort \citep{Oort}). We can take
advantage of this notion to find out useful hints about the population
assignments of the progenitors of novae. Due to their luminosity, M$_V
\gsim -9$ novae are particularly suited for this purpose because they
can be easily identified both in the Milky Way and in external galaxies.

\section{Nova Population in the Milky Way}
Historical data on galactic novae (e.g. McLaughlin
\cite{mcl42},\cite{mcl45},\cite{mcl46} and Payne-Gaposchkin \cite{gapo})
have received discrepant interpretations. Kukarkin \cite{kukas}, Kopylov 
\cite{kopy} and Plaut \cite{plaut} pointed out the existence of a 
concentration of novae towards the galactic plane and the galactic
center and classified them as belonging to the `disk population'.
Minkowski\cite{minko48}, \cite{minko50} and \cite{gapo} showed that
the galactic longitudes of novae and planetary nebulae (PNe) have
similar distributions and therefore novae, like PNe, belong to Pop II
stellar population. Baade
\cite{baa58} assigned novae to Pop II stellar population because of
the occurrence of a few ones (e.g. T Sco 1860) in very old stellar
population systems, such as the Globular Clusters. Iwanowska and
Burnicki \cite{iwa} suggested that novae are a mixture of Pop I and
Pop II objects, and Patterson \cite{pat} proposed that novae belong to
an `old disk' population. Tomaney and Shafter \cite{tom} found that
novae belonging to the bulge of M31 are spectroscopically different
from novae observed in the neighborhood of the Sun and deduced that
galactic novae are mainly `disk' objects. Different 
conclusions were drawn by Della Valle and Duerbeck \cite{dvdu} who
compared the cumulative distributions of the rates of decline for M31,
LMC and Milky Way nova populations (see Fig. 1) and found that
galactic and M31 distributions are indistinguishable, whereas
M31 and LMC distributions are different at $\gsim 99\%$
significance level. Since the speed class of a nova depends on the
mass of the underlying white dwarf (e.g. \cite{shara81}),
systematic differences in the distributions of the rates of decline indicate the
existence of physical differences between LMC and M31 nova
populations. As most novae in M31 are produced in the bulge
(Ciardullo et al. \cite{ciardu}, Capaccioli et al.\cite{capa},
Shafter and Irby\cite{shaf01}), one would argue that novae in the 
Milky Way are also mostly bulge objects.

\begin{figure}
  \rotatebox{0}{\resizebox{20pc}{!}{\includegraphics{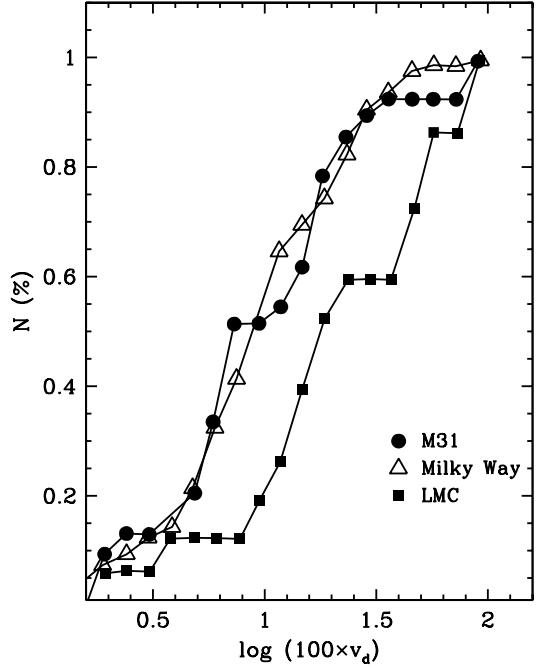}}}
\caption{Cumulative distributions of the rates of decline for M31,  
LMC and Galaxy (from\cite{dvdu})}
\end{figure}

\subsection{Disk and Bulge Novae}

The quantitative characterization of the concept of nova populations
into two classes of objects, i.e. fast and bright `disk novae' and
slow and faint `bulge novae', has been elaborated in the early 90's by
Duerbeck \cite{duer90} and Della Valle et al.\cite{dv92},\cite{dv94},
\cite{dv95},\cite{dv98}. The
former demonstrated that nova counts in the Milky Way do not follow an
unique distribution (Fig. 2), the latter authors showed that the rate
of decline (which traces the mass of the WD associated with the nova
system) correlates with the spatial distributions of the novae inside
the Milky Way (Fig. 3 and 4). Fig. 2 shows that nova counts follow two
different trends. Dashed and dotted lines are the predictions from
simple disk, $\rho(z)=\alpha \times \rho_\circ
\exp(-|z|/z_\circ)$, and bulge ($\rho \sim 10^{0.6}$) nova population
models [with $\rho_\circ =125{\rm pc}$ A$_V=1 {\rm mag/kpc}^{-1}$,
M$_V$(max)$= -9$, $\rho_\circ =10^{-10}$ pc$^{-3}$ yr$^{-1}$ and
$\alpha \lsim 0.1$, see \cite{pat}, \cite{dvdu}, \cite{duer84} 
\cite{nay}]. 

\begin{figure}
\rotatebox{-90}{\resizebox{20pc}{!}{\includegraphics{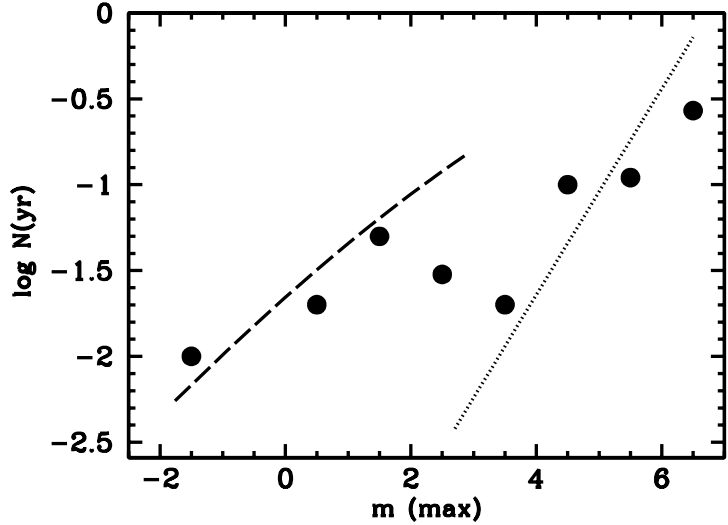}}}
\caption{Theoretical and observed number of novae. Filled circles: observed
rate; dashed line: calculated number counts for a disk population; dotted line:
expected slope for a bulge population. Adapted from \cite{duer90}.}
\end{figure}

A simple consequence of two population distributions is that the rate
of decline is expected to correlate with the galactic
longitude. Fig. 3 compares the distributions of t3 for galactic novae
in the direction of the galactic anti-center ($+90^\circ < l <
270^\circ$, dotted region) and center ($-90^\circ < l < +90^\circ
$). The former distribution mainly formed by `disk' novae, peaks at
t3$\lsim 20^d$, while the latter one peaks at larger t3 because of the
contribution of {\sl slow} `bulge' novae which are viewed in the
direction of the galactic center.

\begin{figure}
\rotatebox{-90}{\resizebox{20pc}{!}{\includegraphics{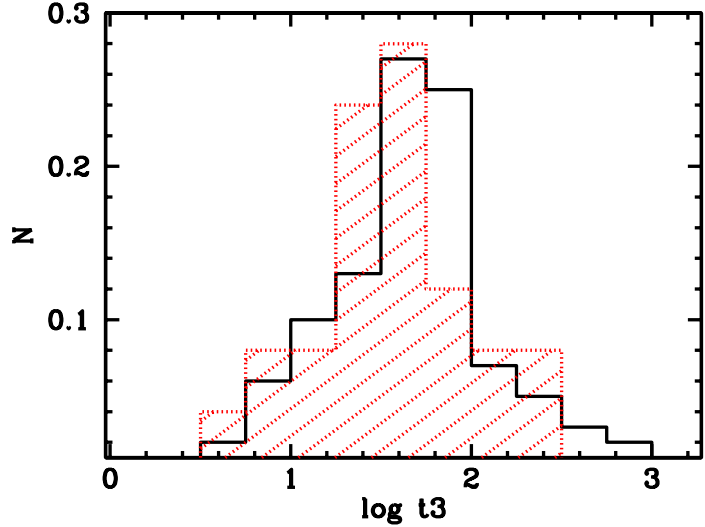}}}
\caption{The distribution of t3 (the time that the nova takes to fall 3 
magnitudes below maximum light) in the direction of the galactic anticenter
($+90^\circ  <l < 270^\circ $, shaded region) and center ($-90^\circ < l <+90^\circ $,
black). Adapted from \cite{dv92}}
\end{figure}

\subsection{The relationship between the rate of decline and height above the 
galactic plane for classical novae}

If galactic novae originate from both the bulge and thin disk of the
Milky Way, then the masses of their WD progenitors are expected to have
heights above the galactic plane systematically different because of
the initial-mass/final-mass relationship for WDs (e.g.\cite{wei}).
This fact can be verified in the following way. Theoretical
calculations (\cite{star85}, \cite{kov}, \cite{kato}) have
established that the strength of the nova outburst is a strong
function of the mass of the underlying WD. On the other hand, the
luminosity of the nova at maximum (which is representative of the
strength of the outburst) correlates with the rate of decline
(\cite{zwi}, \cite{mcl39}, \cite{dvli}). Thus, the distribution of
the rates of decline traces the distribution of the masses of the WDs
associated with the nova systems. We note that the distance moduli
introduced into $z=10^{[0.2(m-M+5-A)]}\times {\rm sin} ~b$ (where $b$
is the galactic latitude) are derived via expansion parallaxes rather
than with the maximum magnitude vs. rate of decline (=MMRD) relationship to
avoid circular arguments. For this reason Della Valle et
al. \cite{dv92} restricted their analysis to a fiducial sample of
only 19 objects coming from Cohen \& Rosenthal
\cite{co83} and Cohen \cite{co85} out of about a hundred novae for which the 
rates of decline have been measured with reasonable accuracy. In this
presentation we were able to augment the original sample by 11 new
objects from \cite{sla}, \cite{dow} and \cite{ring}. Fig. 4 shows
the existence (at a confidence level of $\lsim 3\sigma$) of a
relationship between the rate of decline and the height of the novae
above the galactic plane, ${\rm log}~z=-0.58(\pm 0.18)\times {\rm
(log~100\times v_d)} +2.7(\pm 0.2)$, which can be expressed as ${\rm
log}~z=-1.9(\pm 0.5)\times M_{WD}/M_\odot +3.7(\pm0.6)$, after
recalling that $M_B\propto{\rm log}~M_{WD}$ (from
\cite{livio}) and $M_B\propto {\rm log}~t2$ (from \cite{dvli}).
The trends illustrated in Fig. 4 and 5 indicate that the fastest nova
systems, which contain the most massive WDs, are concentrated close to
the galactic disk. It is difficult to understand how such
distributions could be due to selection effects, since there is no
obvious mechanism to prevent the discovery of fast/bright novae at
high $z$ (although we cannot exclude that some slow/faint nova at
small $z$ can be heavily absorbed and overlooked). Fig. 6 reports the
frequency distribution of the rates of decline (in terms of log t2)
for the fiducial sample of novae. The distribution is bimodal and
shows that `disk' novae, i.e. objects characterized by $z\lsim 150$pc
(shaded region), are mostly `fast' whereas `bulge' novae are mostly
`slow'. The former distribution peaks at M$_V=-8.7^{-0.3}_{+0.6}$ and
the latter at M$_V=-7.2^{-0.9}_{+1.2}$. The existence of such a
bimodality was pointed out by Arp \cite{arp} also for M31 novae (see
his Fig. 36), but this fact was neglected afterwards.

\begin{figure}
\rotatebox{-90}{\resizebox{20pc}{!}{\includegraphics{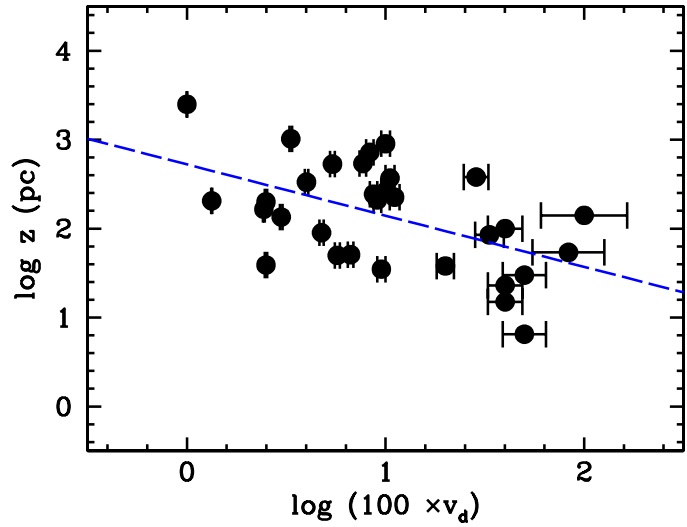}}}
\caption{The relationship between height above the galactic plane vs. rate of 
decline, $v_d=2/t2$. t2 is the time that the nova takes to decrease its
brightness by 2 magnitudes from maximum. The sizes of the errorbars in 
$log~z$ axis are comparable to the sizes of the dots.}
\end{figure}

\begin{figure}
\rotatebox{-90}{\resizebox{20pc}{!}{\includegraphics{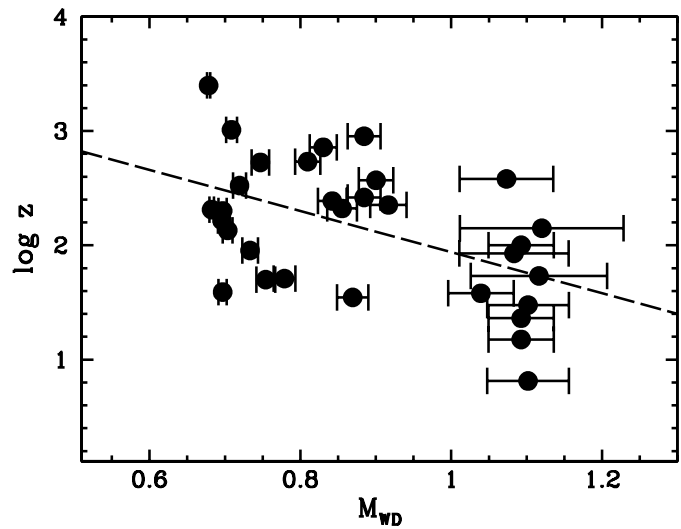}}}
\caption{The relationship between height above the galactic plane vs. mass of 
the underlying WD (in solar masses). The sizes of the errorbars in 
$log~z$ axis are comparable to the sizes of the dots.}
\end{figure}

\subsection{The spectroscopic differences between disk 
and thick-disk/bulge Novae}

Williams \cite{bob} after studying about two dozen of galactic novae
(and a few objects belonging to the LMC) concluded that novae can be
broadly divided into two spectroscopic classes of objects: the Fe II
and He/N novae. The former are characterized by slow spectroscopic
evolution with expansion velocities $\lsim 2500$ km/s (FWZI) and the
Fe II lines as the strongest non-Balmer lines in the early emission
spectrum. The latter are fast spectroscopically evolving novae,
characterized by high expansion velocity ejecta $\gsim 2500$ km/s
(FWZI) with He and N lines being the strongest non Balmer lines in the
emission spectrum near maximum. Hybrid objects (e.g.  V1500 Cyg) that
evolve from Fe II to He/N are classified as FeII-b (b=broad) and are
physically related to the He/N rather than to FeII class. In Fig. 7 we
have plotted the frequency distribution of the heights above the
galactic plane of the novae of the fiducial sample after being
classified according the Williams' criteria. The top and bottom panels
give the distributions of novae which have been classified as He/N
(+FeII-b) and Fe II. The histograms show that novae belonging to the
He/N class tend to concentrate close to the Galactic plane with a
typical scale height $\lsim 150$ pc, whereas FeII novae are
distributed more homogeneously up to $z \sim 1000$ pc and beyond. A
K-S test on the data shows that the two distributions are different at
$\gsim 95$\% level.  Fig. 7 indicates that the objects previously
classified as `disk' and `bulge/thick-disk' novae tend to correspond
to the spectroscopic classes introduced by Williams, indeed about 70\%
of fast and bright novae belong to the He/N (or FeII-b) class, while
the slow and faint ones form the main bulk of Fe II class. To explain
this behavior one should consider the following. The more massive the
WD (for a given $\dot M$ and T$_{WD}$) the smaller is the mass of the
accreted envelope (in view of $\Delta M_{acc} \propto
R^4_{WD}/M_{WD}$) the more violent is the outburst (i.e. shorter t3
and higher expansion velocities) and the larger the fraction $\Delta
m_{shock}/\Delta m_{wind}$ where $\Delta m_{shock}$ is the shell mass
ejected at the maximum and $\Delta m_{wind}$ is the fraction of the
shell mass ejected in the subsequent continuous optically-thick wind
phase. Since He/N spectra are formed in the shell ejected at the
outburst maximum (as one can infer from the top-flatted profiles of
the emission lines, for example) it is very likely that He/N novae are
generally associated with massive WDs. If this interpretation is
correct the distributions reported in Fig. 7 are simple consequences
of the trends reported in Fig. 4 and 5.

\begin{figure}
\rotatebox{-90}{\resizebox{20pc}{!}{\includegraphics{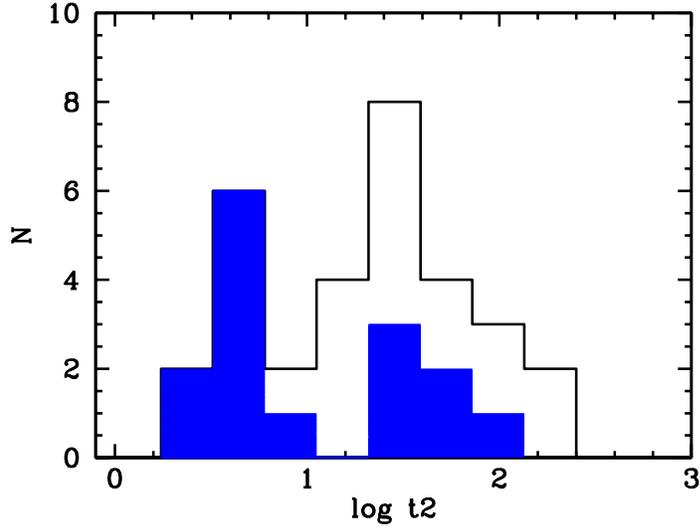}}}
\caption{Frequency distribution of the rates of decline for the novae
of the fiducial sample. The shaded region represents novae with $z
\lsim 150$pc.}
\end{figure}

\begin{figure}
\rotatebox{-90}{\resizebox{20pc}{!}{\includegraphics{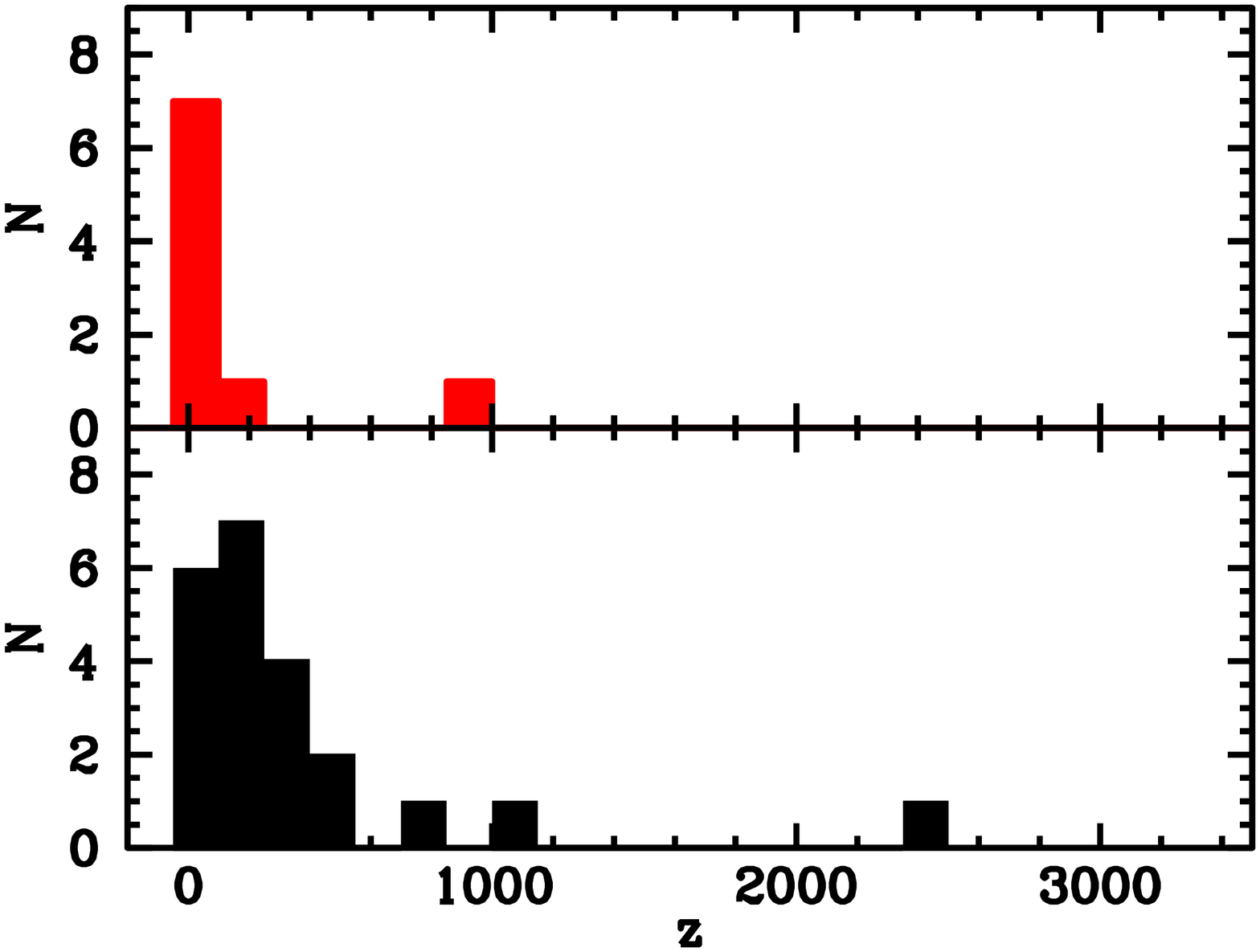}}}
\caption{Frequency distribution of the height above the galactic plane for 
He/N (and FeII-b) novae (top panel) and FeII novae (bottom panel). Adapted
from [26].}
\end{figure}

\section{Nova populations in external galaxies} 

The use of galactic data to establish the stellar population where
novae originate has been often questioned because of observational
bias, mainly due to both interstellar absorption in the galactic disk
and our position within the Galaxy. These effects can be largely
minimized by studying the nova populations in external galaxies,
particularly: 1) their spatial distribution; b) the maximum magnitude
vs. rate of decline relationship; c) differences (if any) in the nova
rates exhibited by galaxies of different Hubble type. The available
data are summarized in Table. 1. Col. 1 gives the galaxy
identification; col. 2 the nova rate; col. 3 the total B mag of the
galaxy corrected for background and foreground absorption; col. 4 the
color (B--K); col. 5 the adopted distance modulus; col. 6 the
normalized nova rate, i.e. the nova rate per year/$10^{10}L_{K\odot}$;
col. 7 the Hubble type. Data come from \cite{dv94} and \cite{sha00}.
Note that to be consistent with the results of MACHO\cite{wel} and
EROS\cite{gli} teams, who found 4 novae in the SMC after one year of
monitoring to search for microlensing events, the nova rate in this
system has been estimated, on the basis of simple statistical
arguments, to be $\sim$ twice as large as quoted in \cite{sha00}.

\begin{table}
\caption{Absolute and normalized nova rates}
\begin{tabular}{ccccccc}
\hline
       &           &         &       &               &        &            \\
Galaxy &Novae/yr   &B$_{tot}$& (B--K) & (m--M)        & $\nu_K$ & T         \\ 
       &            &         &       &              &                 \\
LMC    &$2.5\pm 0.5$& 0.57    & 2.74  &$18.58\pm 0.1$ &  $5.1\pm 1$  &9  \\
SMC    &$0.7\pm 0.2$& 2.28    & 2.71  &$19.00\pm 0.1$ &  $4.8\pm 1.5$&9  \\
M33    &$4.6\pm 0.9$& 5.75    & 2.87  &$24.64\pm 0.2$ &  $3.7\pm 0.9$&6  \\
M101   &$12\pm 4$   & 8.26    & 3.24  &$29.35\pm 0.2$ &  $0.9\pm 0.3$&6  \\
M51    &$18\pm 7$   & 8.41    & 3.43  &$29.60\pm 0.2$ &  $1\pm 0.4$  &4  \\
M31    &$29\pm 4$   & 3.51    & 3.85  &$24.42\pm 0.2$ &  $1.5\pm 0.2$&3  \\
M81    &$24\pm 8$   & 7.39    & 3.99  &$27.80\pm 0.2$ &  $1.7\pm 0.6$&2  \\
N5128  &$28\pm 7$   & 6.32    & 3.38  &$27.80\pm 0.2$ &  $3.3\pm 0.8$&--2  \\
N1316  &$130\pm 40$ & 9.20    & 4.15  &$31.50\pm 0.3$ &  $1.7\pm 0.5$&--2  \\
M87    &$91\pm 34$  & 9.49    & 4.17  &$30.90\pm 0.2$ &  $2.1\pm 0.8$&--4  \\
VirgoEs&$160\pm 57$ & 9.46    & 4.26  &$31.35\pm 0.2$ &  $2.2\pm 0.8$&--4  \\
       &            &         &       &               &                 \\
\hline
\end{tabular}
\end{table}

\subsection{The Spatial Distribution}

Due to its closeness, the M31 nova population has been the only one
extensively studied. However a simple skimming through past literature
reveals the existence of different ideas on its nova population
assignment: disk populations (Arp \cite{arp}, Rosino \cite
{leo});halo population (Wenzel and Meinunger \cite{wenz}); mainly
bulge population (Ciardullo et al. \cite{ciardu}, Capaccioli et
al. \cite{capa} see their Fig. 7); mostly disk population (Hatano et al. \cite{hat}). 
\medskip

It is very likely that a fraction of M31 novae considered by
\cite{ciardu} and \cite{capa} as belonging to the bulge are in fact
(according to \cite{hat} see their Fig. 1 and 4) physically related to
the disk and their allotment to the bulge was a simple consequence of
neglecting the geometrical projection effect. However Shafter \& Irby
\cite{shaf01} have reappraised the Hatano et al's
conclusions, after comparing the spatial distribution of novae with
the background M31 light (see their Fig. 4). They confirm most M31 novae to be originated
in the bulge (about $70\%$ up to a minimum of 50\%).  Concerning this
last point a word of caution seems in order as long as the effects of
extinction on nova detections in the M31 disk will be fully
quantified.

M33 is an almost bulgeless galaxy \cite{bot} and therefore its nova
production necessarily originates in the disk. Particularly Fig. 2 of 
\cite{dv94} shows that most novae appear superimposed on the arms of the 
parent galaxy. Similar arguments hold for the LMC.  

The analysis of 78 plates obtained in the 1950-55 Palomar campaign for
the discovery of novae in M81 has been recently published by Shara,
Sandage and Zurek \cite{mike}. These authors find that the spatial
distribution of novae in M81 is fully consistent with the two nova
populations hypothesis (see their Fig. 1). Particularly they estimate
that the fraction of novae belonging to the M81 bulge is not larger than
$\sim 60\%$.

\subsection{The Maximum Magnitude vs. Rate of Decline Relationship} 

Fig.8 reports the maximum magnitude vs. rate of decline relationship
for LMC, M31 and Virgo novae.  A simple glance indicates that the nova
production in the LMC is clearly biased towards fast and bright novae,
whereas the M31 nova population exhibits a prominent `slow'
component. By the same token, admittedly on the basis of a scanty
statistic, the trend exhibited by novae discovered in Virgo (Pritchet
\& van den Bergh \cite{vdb}). It is not obvious to explain this
behavior in terms of an observational bias: indeed the brightest novae
are detected in the nearest galaxy (LMC) and would be missed in the
more distant ones (Virgo). On the other hand, the differences in the
MMRDs find a simple explanation in the framework of the two nova
populations scenario: nova systems in disk dominated galaxies are
associated with more massive WDs, then resulting in faster and
intrinsically brighter nova events.

\begin{figure}
\rotatebox{-90}{\resizebox{25pc}{!}{\includegraphics{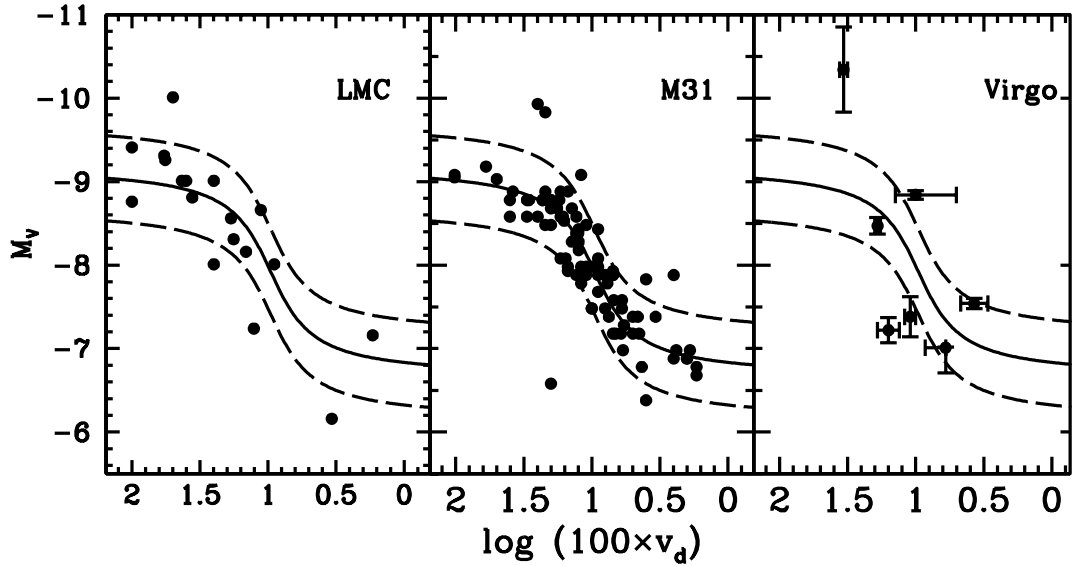}}}
\caption{The MMRD relationship for LMC, M31 and Virgo nova populations.
The solid line indicates the best fit (from \cite{dvli}). Upper and lower
dashed curves are located at $\pm 3\sigma$.}
\end{figure}

\subsection{Is the nova rate depending on the Hubble type of the parent 
galaxy?}

Della Valle et al. \cite{dv94} suggested that there may exist a
systematic difference in the nova rate per unit of K (and H)
luminosity in galaxies of different Hubble types (see their Fig. 3 and
4). These authors found that galaxies of late Hubble types are more
prolific nova producers than early-type ones by a factor $\sim
3$. This overproduction of novae could be the consequence of
``selection effects'' on the nova frequency (Truran \& Livio
\cite{tru86}, Ritter et al. \cite{rit}). For example, if we use the
results of \cite{rit}, we obtain for the ratio of the nova outbursts
density, $\rho(disk)_{outburst}/\rho(bulge)_{outburst}\sim 4$ in good
agreement with the observations. A possible interpretation is that
nova systems in disk dominated galaxies result in {\underline {more
frequent nova events}} due to the shorter nova recurrence time
associated with massive WDs (see \cite{tru90}). However this finding
has been put into question by \cite{sha00} (see also \cite{sharov})
on the basis of nova rates on M51 and M101. Fig. 9 shows the trend of
the normalized nova rate $\nu_K$ vs. (B--K) as derived from
Tab. 1. With the exception of M101 and M51 (for which the nova rates
reported by \cite{sha00} may be lower limits) the normalized nova
rate increases from $\nu_K \lsim 2$ novae yr$^{-1}/10^{10} L_{K\odot}$
for early types to about $\nu_K \sim 5$ for late types.

\begin{figure}
\rotatebox{-90}{\resizebox{20pc}{!}{\includegraphics{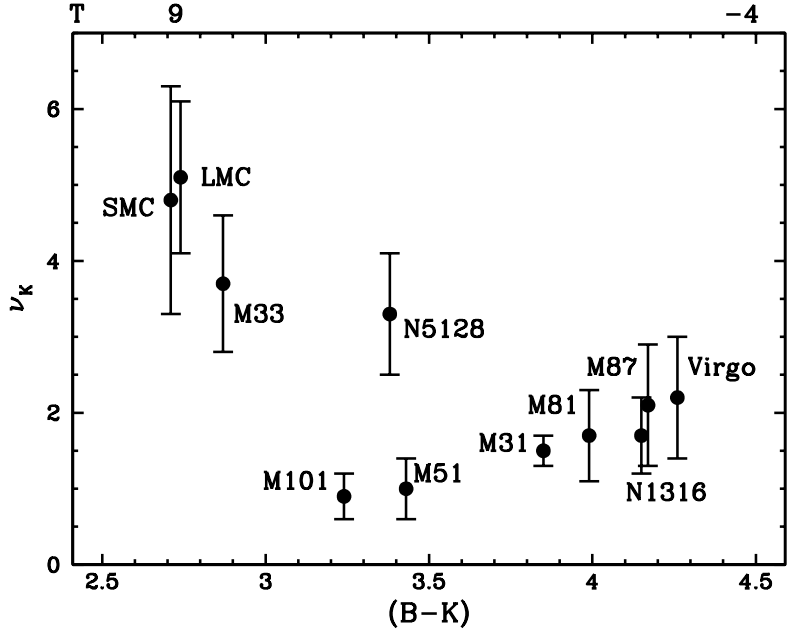}}}
\caption{The nova rate per unit of K luminosity as a function of the Hubble 
type of the parent galaxy. The error bars reflect only the poissonian errors
affecting the single nova rates.}
\end{figure}

\section{Conclusions}

Both observations of galactic and extragalactic novae converge toward
the idea that novae are drawn from two different stellar
populations. In particular:

1) analysis of the nova counts and rates of decline inside the Milky
Way has allowed to derive the notion of disk and bulge/thick disk nova
populations. The typical {\sl disk nova} is a fast evolving object
whose lightcurve exhibits a bright peak at maximum M$_V\gsim -9$
(t2$\lsim 13^d$ or t3$\lsim 20^d$) a smooth early decline and belongs
to the He/N (or Fe IIb) spectroscopic class. The progenitor is
preferentially located at small heights above the galactic plane
($\lsim 150$pc) and since it is related to (a relatively old) Pop I
stellar population, the associate WD is rather massive, M$_{WD}\gsim
1$ M$_\odot$. The typical bulge/thick disk nova
is a slow evolving object whose lightcurve exhibits a fainter peak at
maximum, M$_V\sim -7.2$ (t2$\gsim 13^d$ or t3$\gsim 20^d$), often
double maxima, dust formation, maximum standstill and belongs to the
Fe II spectroscopic class. The progenitors extend up to $1000$ pc from
the galactic plane and are likely to be related to a Pop II stellar
population of the galactic thick-disk/bulge and are therefore
associated (on average) with less massive WDs, M$_{WD}\lsim 1$M$_\odot$.

2) analysis of the MMRD relationship for LMC, M31 and Virgo novae
confirm the existence of systematic differences in the
distributions of the rates of decline of the respective nova
populations. 80\% of novae in the LMC are bright and fast and
therefore associated (on average) with massive WDs while fast novae
in M31 are $\lsim 25$\%. The two distributions are significantly different
(a K-S test gives $\gsim 99$\%). 
The analysis of the spatial distributions suggests that nova
populations in M31 and M81 are a mixture of disk and bulge novae (
$\sim 30\%-70\%$ and $40\%-60\%$ respectively). Novae in the LMC and M33
originate from disk population whereas novae in M87 and
NGC 1316 are related `a fortiori' to bulge population.

3) the previous results imply the existence of differences in the age of the
progenitors of disk and bulge novae. This has been recently proven
by Subramanian \& Anupama \cite{anup}. These authors
have studied the star formation history of nearby regions around LMC
novae and show that most of them occur in
the bar or close to it. They conclude that the parent stellar
population of the fast and slow novae are likely to be in the range
$\lsim 1-3$ and $\gsim 3-10$ Gyr respectively.

4) some of the differences between disk and bulge novae, as described above, 
are expected on the basis of theoretical arguments (see Kolb \cite{kolb},
Starrfield et al. \cite{star98}, Kato \cite{kato2}).

5) Analysis of the distributions of the rates of decline as a function of
the absolute magnitude at maximum of LMC, M31 and Virgo, reveals the
existence of a scant group ($\lsim 5\%$) of super-bright novae which
deviate systematically from the MMRD relationship by more than one
magnitude \cite{dv91}.  One possible explanation is that a
{\sl ``super-nova''} explosion might occur at the end of the life of a CV
(\cite{iben}, see also\cite{iben2}). 
The recent study on Nova LMC 1991 (\cite{star01})
suggests that the metallicity may be the driving parameter to account for
this deviating behavior.

\subsection{Future studies}

Whether or not the nova rate depends on the Hubble type of parent
galaxy is still an open question. If low nova rates for M51 and M101
will be confirmed a possible explanation for the trend reported in
Fig. 9 has been pointed out by Yungelson, Livio and Tutukov
\cite{lev} (see also \cite{tut}). These authors noted that small
$\nu_k$ are associated with high-mass spirals such as M51 and M101,
while high values of $\nu_k$ are typical of low mass galaxies such as
LMC and M33.  If the presently observed rates of disk novae in late
type galaxies are mainly determined by the current SFRs, one expects
to observe differences in the normalized rates of high- and low-mass
spirals. Indeed for the former the SFR in the past was several ($\lsim
5$) times higher than the present one (Gallagher et
al. \cite{gal}). For the latter the SFR was nearly constant over the
galaxy lifetime (e.g. Gavazzi \& Scodeggio \cite{gav}). Therefore
high-mass spirals contain a higher fraction of old red stars than
low-mass ones and consequently also their IR-luminosity is higher
(this is actually observed, see col. 4 in Tab.1).  As a consequence,
since the K luminosity is proportional to the mass in old stars, one
may expect $\nu_k$ in systems such as M33 or LMC to be higher than
that in M51 or M101. It is apparent that this issue will be solved as
soon as new nova rates in external galaxies will be made
available. For example, preliminary results, based on HST observations
(e.g. \cite{mike2}), may suggest high values of nova rates also in
elliptical galaxies. We note that nova survey in extragalactic
systems, despite their scientific interest, have not been popular
among astronomers (although `remarkable' exceptions
do exist, e.g. \cite{hub}). Probably the main reason for this is the
unpredictable nature of nova events, which made nova surveys
considerably (telescope) time consuming. However, the coming into
operations of 8-10m class telescopes should change this bias. Della
Valle and Gilmozzi \cite{dv02} have carried out with VLT a pilot
programme to discover novae in NGC 1316.  They found 4 novae with 3h
of observing time (about $\sim 0.8^h$ per nova). A similar programme
carried out by \cite{vdb} with a 4m class telescope on Virgo galaxies
(which should have comparable nova rates and distance than NGC 1316)
discovered 9 novae in 56h ($\sim 6.5^h$ per nova). This experiment
teaches us that the 8-10m class telescopes equipped with larger and
more efficient detectors are able to improve the payoff per night, in
terms of nova detections in galaxies outside the Local Group, by on
order of magnitude with respect to the previous generation of
telescopes.


\begin{theacknowledgments}

I'm grateful to Nino Panagia and Bob Williams for useful comments.

\end{theacknowledgments}




\end{document}